\newif\ifarxiv %
\arxivtrue  %

\ifarxiv %

    \documentclass{article}
    \usepackage{arxiv}
    \usepackage{multirow}
    \usepackage[utf8]{inputenc} %
    \usepackage[T1]{fontenc}    %
    \usepackage{hyperref}       %
    \usepackage{url}            %
    \usepackage{booktabs}       %
    \usepackage{amsfonts}       %
    \usepackage{nicefrac}       %
    \usepackage{microtype}      %
    \usepackage{doi}

    \newcommand{\approach}{Misunderstanding to Mastery\xspace}
    \newcommand{\appAbb}{M2M\xspace}
    
    \usepackage{colortbl}
    \usepackage{subcaption}
    \usepackage{graphicx} %
    \usepackage[numbers]{natbib} 
    \usepackage{xcolor}
    \usepackage{tabu}
    \usepackage{xspace}
    \usepackage{listings}
    \usepackage{float} 
    \usepackage[symbol]{footmisc}
    
    \usepackage{algorithm}
    \usepackage{algpseudocode}
    \usepackage{amsmath}
    \usepackage{amssymb}
    
    \title{From Misunderstandings to Learning Opportunities: Leveraging Generative AI in Discussion Forums to Support Student Learning}  

    \author{
        Stanislav Pozdniakov \\
        School of Electrical Engineering and Computer Science \\
        The University of Queensland \\
        St Lucia, QLD, 4072, Australia \\
	\texttt{s.pozdniakov@uq.edu.au} \\
    \And
        Jonathan Brazil \\
        Institute for Teaching and Learning Innovation \\
        The University of Queensland \\
        St Lucia, QLD, 4072, Australia \\
        \texttt{j.brazil@uq.edu.au} \\
    \And
        Oleksandra Poquet \\
        School of Social Sciences and Technology \\
        Technical University of Munich \\ 
        Munich, Germany \\
        \texttt{sasha.poquet@tum.de} \\
    \And
	Stephan	Krusche \\
        Applied Education Technologies \\
        Technical University of Munich \\
        Munich, Germany \\
	\texttt{krusche@tum.de} \\
    \And
        Santiago Berrezueta-Guzman \\
        Technical University of Munich \\
        Munich, Germany \\
        \texttt{s.berrezueta@tum.de}
    \And
	Shazia Sadiq \\
        School of Electrical Engineering and Computer Science \\
        The University of Queensland \\
        St Lucia, QLD, 4072, Australia \\
	\texttt{shazia@eecs.uq.edu.au } \\
    \And
        Hassan Khosravi \\
        Institute for Teaching and Learning Innovation \\
        The University of Queensland \\
        St Lucia, QLD, 4072, Australia \\
	\texttt{h.khosravi@uq.edu.au} \\
}
    \renewcommand{\shorttitle}{M2M: Transforming Student Dialogues into Learning Opportunities}

\begin{document}
    \maketitle
    \begin{abstract}
In the contemporary educational landscape, particularly in large classroom settings, discussion forums have become a crucial tool for promoting interaction and addressing student queries. These forums foster a collaborative learning environment where students engage with both the teaching team and their peers. However, the sheer volume of content generated in these forums poses two significant interconnected challenges: How can we effectively identify common misunderstandings that arise in student discussions? And once identified, how can instructors use these insights to address them effectively? This paper explores the approach to integrating large language models (LLMs) and Retrieval-Augmented Generation (RAG) to tackle these challenges. We then demonstrate the approach \approach (\appAbb) with authentic data from three computer science courses, involving 1355 students with 2878 unique posts, followed by an evaluation with five instructors teaching these courses. Results show that instructors found the approach promising and valuable for teaching, effectively identifying misunderstandings and generating actionable insights. Instructors highlighted the need for more fine-grained groupings, clearer metrics, validation of the created resources, and ethical considerations around data anonymity.

\end{abstract}

    
    \section{Introduction}  
\label{sec:intro}

Discussion forums are a pedagogically robust medium for fostering student engagement in asynchronous conversations, providing significant benefits, including flexible participation, access to varied viewpoints, and deeper contemplation of course material \cite{almatrafi2018systematic}. However, the volume of content generated by students, often in large foundational courses, poses notable challenges. As students contribute to and engage with numerous lengthy threads, it becomes challenging for instructors to efficiently identify and address key misunderstandings \cite{alrajhi_multidimensional_20,yang_untangling_22}. Automating the process of pinpointing these misunderstandings can help reduce instructors' cognitive load, allowing them to focus their expertise on evaluating and refining the most critical issues and provide insight on what areas need to be clarified through teaching to bring more students along, significantly enhancing student learning. 

Given their ability to process extensive text, LLMs could effectively detect misunderstanding patterns. Integrating RAG, which combines information retrieval with LLM generation capabilities, can overcome limitations when applying LLMs in disciplinary contexts \cite{gao_rag_24}, demonstrating the technique's potential to deliver contextually relevant insights \cite{taneja_jill_24,yan_vizchat_24,lin_improving_24}. However, insights alone are insufficient; as shown in learning analytics dashboard studies, without technological tools to operationalize them, these insights often remain unused \cite{kaliisa2024have}. Therefore, a robust technological solution is needed to identify and effectively address misunderstandings.

This work introduces the approach \approach (\appAbb), leveraging LLMs and RAG to identify and address student misunderstandings in large courses, while maintaining instructor oversight. We demonstrate its application in three computer science courses and evaluate it with five instructors via structured interviews.

    \section{Related work}
\label{sec:related-works}

\textbf{Identifying student misunderstandings.} Early approaches to identify student misunderstandings, such as quizzes with closed- or open-ended questions, often require human labelling and lack the responsiveness required to offer students timely feedback. A semi-automated method requiring minimal human labelling to predict misunderstandings through statistical co-occurrence patterns was suggested by \cite{stephens2016identifying}.  Recent advances in Knowledge Tracing (KT) enable real-time detection by moving beyond binary assessments \cite{park_CBKT_24}, but these models are restricted to MCQs and depend on ad hoc mappings between concepts and misunderstandings. While data-driven methods have advanced, no prior work has used LLMs to identify misunderstandings from discussion forums -- a key contribution of our paper.

\textbf{Automatic generation of learning resources.} Previous research has shown a strong interest in exploring how LLMs can be leveraged for the automatic generation of learning resources. For example, \citep{denny_CanWeTrust_23} examined how LLMs could generate educational resources comparable to those produced by computer science students, finding that LLM-generated resources were less variable in length and more consistent with existing examples. Other research has focused on leveraging LLMs for the creation of specific types of learning resources such as worked \citep{jury_evalLLMWorked_24}, MCQs \citep{hwang_mcqGenAI_24} and even transforming lecture transcripts into educational dialogues \citep{choi_VIVID_24}.  While evaluation results highlighted LLMs' capabilities, challenges like GenAI hallucinations and shortcomings in the depth and structure of learning resources elements remain. RAG has been proposed as a solution, as it grounds AI outputs in verifiable data, improving accuracy and reducing errors \citep{wang_LLMsED_24,gao_rag_24}. Our approach does not introduce new methods for content creation; rather, it leverages best practices from existing approaches to generate prompts for content creation, ensuring reliable and effective instructional material.

    \section{Approach: \approach (\appAbb)}\label{sec:approach}

\begin{figure*}[htp] %
\centering
\includegraphics[width=.85\textwidth,keepaspectratio]{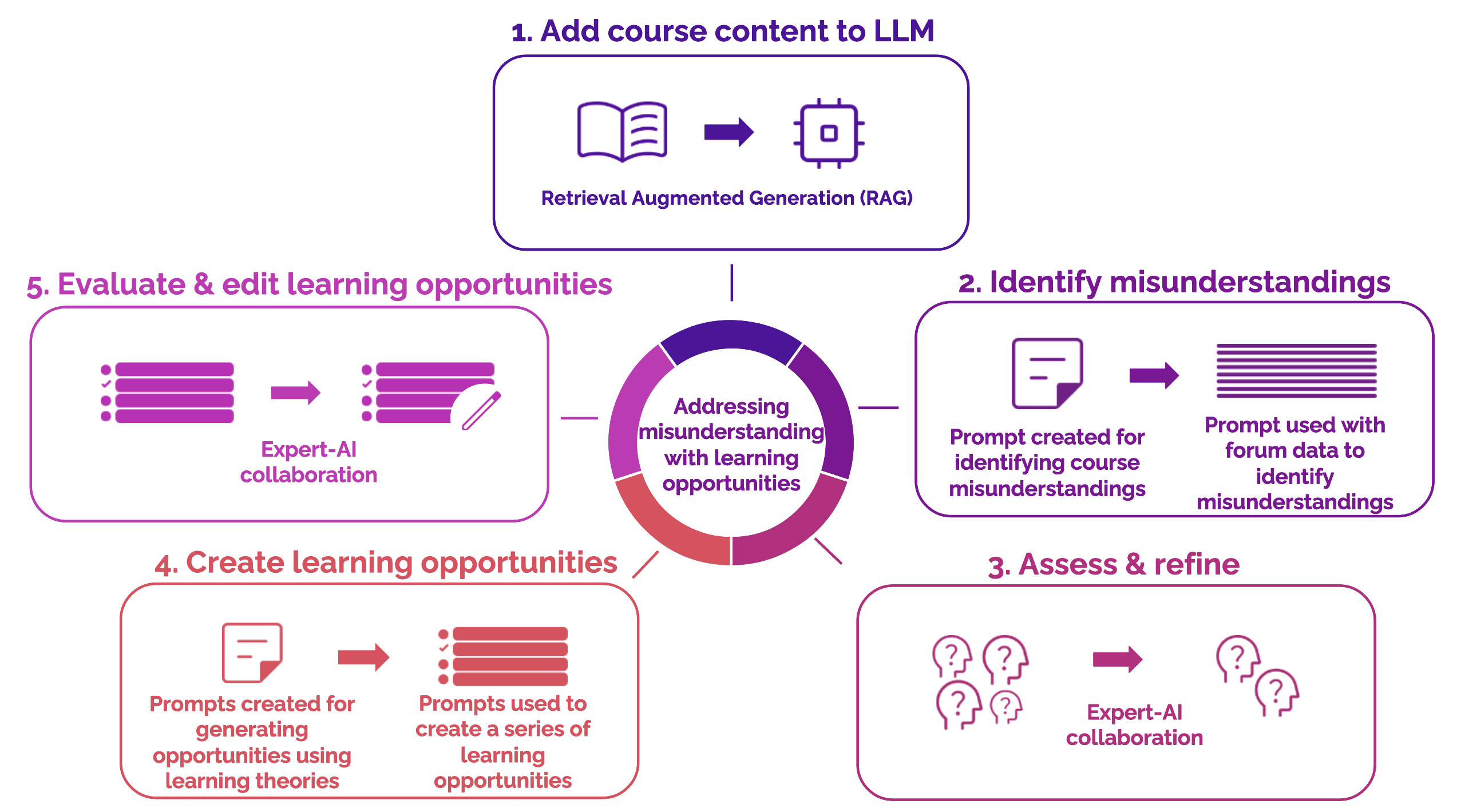}
\caption{
Overview of the proposed approach in \appAbb to discover misunderstandings from students' discussion forums and to guide instructors to use them to create learning opportunities.
}
\label{fig:approach}
\end{figure*}

In this section, we present \appAbb approach, which aims to tackle the dual challenges of identifying common misunderstandings in discussion forums and developing effective learning resources to address them. An overview of the approach is shown in Figure~\ref{fig:approach}. The proposed approach consists of \textbf{five steps}, building upon the authors' prior work in developing a framework for the ethical application of GenAI to support instructional activities \cite{anon_reference_llm_meets} and drawing on the approach outlined in \cite{choi_VIVID_24}. We provide a detailed description of what each step of the approach entails.

\textbf{Step 1: Add course content to the LLM.}
Building upon the increasing evidence of the benefits of RAG  for educational tasks \cite{lin_improving_24,yan_vizchat_24}, we augment the underlying LLM used in \appAbb by incorporating auxiliary course content, including lecture notes, tutorials, and assessment tasks. Technically, this step involves the creation of a vector database where the embedding representations of the learning materials are stored, enabling their use in the RAG architecture with LLMs, development of querying logic for content and the prompts that will guide the LLM's interaction with this content. Conceptually, once the relevant course content is identified, instructors must carefully consider the retrieval instructions from the vector database -- such as the criteria for querying specific learning content.

\textbf{Step 2: Identify Misunderstandings.}  The second step involves developing prompt pipelines that leverage students' discussion forum data to uncover class-level misunderstandings. This process involves developing prompts to analyze students' posts individually or in batches to detect misunderstandings, and summarize and refine them into broader class-level categories. For instructional validity and scrutable review of intermediate results, these pipelines must align with established educational frameworks and pedagogical theories (e.g., \cite{anon_reference_llm_meets,mollick_AssigningAI_23}) or closely replicate the methods used by disciplinary experts. Advanced automatic prompt optimization with LLM fine-tuning could enhance effectiveness, but obtaining a reliable baseline for systematic experimentation remains a challenge. Crucial is LLM selection, balancing time, cost, and task complexity. Smaller, efficient LLMs enable real-time use via speed and lower costs, whereas reasoning LLMs offer higher precision and context at greater cost and slower speeds, potentially needing asynchronous delivery. Balancing these trade-offs and providing instructor options is vital for choosing the optimal approach within resource and pedagogical constraints.

\textbf{Step 3: Assess and refine the identified misunderstandings}. This step involves collaboration between instructors and AI to assess and refine misunderstandings identified in student questions from discussion forums. To support this process, \appAbb computes two key metrics in \textit{coverage}, which is the number of posts related to a specific misunderstanding, and \textit{cohesion}, which measures how semantically consistent and tightly related the posts are to the identified misunderstanding. Coverage is determined by a classification algorithm that associates each post with zero or more identified misunderstandings. Cohesion is calculated by obtaining embeddings of the related posts, computing their centroids, and then measuring the average cosine similarity between the post embeddings and their centroids. %

\textbf{Steps 4 and 5: Create and evaluate targeted learning opportunities.} Step 4 uses RAG with course content and misunderstanding data to generate tailored learning opportunities targeting these misunderstandings. \appAbb employs state-of-the-art LLM prompting techniques \cite{denny_CanWeTrust_23, denny_PromptProblemsNew_24, jury_evalLLMWorked_24, hwang_mcqGenAI_24}, particularly prompt chaining, noted for superior results \cite{jury_evalLLMWorked_24}. This involves modular prompts where one LLM's output feeds into the next, potentially using few-shot examples. The three-part prompt chain uses the misunderstanding description and associated posts as initial input: \textit{idea brainstorming}, resulting in learning resource ideas; \textit{idea selection}, which chooses the best idea and generates creation steps; \textit{learning resource self-refinement}, which retrieves relevant materials using the misunderstanding description to align the opportunity's difficulty and thinking level.
Step 5 involves a collaborative effort between instructors and AI to evaluate and refine the quality of the generated learning opportunities. \appAbb suggests using an LLM to provide initial feedback based on predefined criteria such as correctness, contextual depth, the quality of distractors (for MCQs), and alignment with the identified misunderstanding, helping instructors to view the AI feedback and decide whether to regenerate the opportunity, remove it if it is fundamentally flawed, or manually edit it to improve its quality. %

    \newcommand{\algoStudents}{574\xspace}
\newcommand{\algoPosts}{1800\xspace}
\newcommand{\algoComments}{4700\xspace}

\newcommand{\webSysStudents}{225\xspace}
\newcommand{\webSysPosts}{521\xspace}
\newcommand{\webSysComments}{1200\xspace}

\newcommand{\infSysStudents}{556\xspace} %
\newcommand{\infSysPosts}{557\xspace} %
\newcommand{\infSysComments}{597\xspace} %

\section{Evaluation}
To assess our approach, we implemented it in three CS courses with a large student intake and held structured interviews with five (\textbf{P1-P5}) experienced university instructors teaching these courses. P1, P4, and P5 taught Information Systems (\infSysStudents students, \infSysPosts posts containing \infSysComments comments); P2, taught Web Information Systems (\webSysStudents students, \webSysPosts posts, and \webSysComments comments); and P3 taught Algorithms and Data Structures (\algoStudents students, \algoPosts posts, and \algoComments comments). Our evaluation is guided by the following \textbf{RQ}: \textbf{What are instructors' perspectives on the effectiveness, adoption, and impact on student learning of the \appAbb approach}?

For the implementation, we used OpenAI's GPT-4o mini for discovering misunderstandings and Google's LearnLM-1.5-Pro-Experimental for generating learning resources and providing resource evaluation feedback. Our prompt design combines zero-shot prompting, chaining, and self-refinement techniques\footnote{The interview guide, overview of \appAbb architecture and prompts are available at \url{https://github.com/stlkcmrd/AIED25-M2M}\label{foot:repo}}. To emphasize the role of interfaces and visual guidance while keeping human expertise central to the pedagogical process, the outputs were presented on the low-fidelity prototype.

\textbf{Method.} Two researchers interviewed participants about their teaching experience, GenAI proficiency, and forum use. Participants confirmed frequent forum use for identifying issues, especially near deadlines and in new courses. To address RQ, participants reviewed the \appAbb approach and results via a prototype, discussing conceptual benefits and practical integration. Once the authors implemented the approach in three CS courses (Section~\ref{sec:approach}), participants evaluated AI-identified misunderstandings for authenticity and prevalence and assessed corresponding MCQs for quality and alignment. Interviews were recorded, auto-transcribed via Zoom, and notes were taken. Two qualitative researchers reviewed transcripts and notes, and then grouped responses into broad themes. These responses, framed by RQ, are presented in Table~\ref{tab:eval-results}.

\textbf{Results.} All participants equivocally supported the idea of the identification of misunderstandings and the generation of learning opportunities. Participants indicated they would like to have and use such a tool weekly, with \textbf{P3} suggesting it could reduce the cognitive load associated with creating these learning opportunities. \textbf{P5} emphasized \appAbb's pedagogical value, suggesting it improves teaching quality by enabling more targeted responses, going beyond simple workload reduction. There was a consensus, mentioned by \textbf{P1}, that the tool complements instructor expertise rather than replacing it, making it essential for instructors (\textbf{P1, P3, P5}) to act as a ``final filter'' for the generated outputs. While instructors (\textbf{P2, P5}) believed using \appAbb could positively impact students' learning through targeted support, \textbf{P5} raised the concern of balancing these opportunities effectively without overburdening students, acknowledging diverse student preferences for learning resources.

\begin{table}[H]
\tiny
\centering
\caption{Summary of Key Insights.\label{tab:eval-results}}
\centering
\fontsize{11}{12.64}\selectfont
\resizebox{0.99\columnwidth}{!}{
\begin{tabular}[t]{>{\raggedright\arraybackslash}p{3.4cm}>{\raggedright\arraybackslash}p{16.1cm}}
\toprule
\multicolumn{1}{c}{RQ} & \multicolumn{1}{c}{Main Insights}\\
\midrule
 & \textbf{-} Equivocal support of the idea of identification of misunderstandings and generation of learning opportunities as ``\textbf{insightful}'' and ``\textbf{actionable}'' (P1-P5)\\

 & \textbf{-} An \textbf{additional step} could be required to evaluate the effectiveness of learning opportunities but might be difficult to implement (P5).\\

\multirow{-3}{3.4cm}{\raggedright\arraybackslash \textbf{\appAbb approach}} & \textbf{-} Instructors would benefit from a \textbf{temporal presentation} of misunderstandings, such as weekly or bi-weekly filtering (P3\&P5).\\
\cmidrule{1-2}
 & \textbf{-} Identified misunderstandings were genuine and common among students, and learning opportunities were well-received.\\

\multirow{-2}{3.4cm}{\raggedright\arraybackslash \textbf{Authenticity \& Effectiveness}} & \textbf{-} Adding ``\textbf{severity ranking}'' and ``\textbf{concept weightiness}'' could additionally aid instructors in assessing the criticality of misunderstandings.\\
\cmidrule{1-2}
 & \textbf{-} \appAbb could support instructors in a variety of \textbf{pedagogical tasks}, i.e., coordinating efforts among instructors, targeting responses, and improving course design, by aiding their awareness and enabling structural reflection on students' knowledge gaps. (P1-P5).\\

 & \textbf{-} Instructors saw themselves as a ``\textbf{final filter}'' before presenting insights from \appAbb to students (P1, P3, P5).\\

\multirow{-3}{3.4cm}{\raggedright\arraybackslash \textbf{Implications for Instruction \& Learning}} & \textbf{-} \appAbb could impact students by reassuring them \textbf{they are not alone} in their struggles (P2, P4), but there is a \textbf{risk of overwhelming them with the additional workload}.\\
\bottomrule
\end{tabular}
}
\end{table}

\section{Conclusion}
This paper introduces an approach which leverages LLMs and RAG to identify class misunderstandings and generate targeted activities. Evaluated with five instructors using data from three CS courses, the approach shows GenAI's potential for educational applications. Pedagogically, it enables instructors and AI to collaborate in detecting hidden learning gaps, beyond just optimizing workload. Technologically, it integrates LLMs into educational tools, an emerging area \cite{anon_reference_llm_meets, glassman_AIResilientUIs_24}. Socially, it supports rather than replaces collaborative learning. While instructors see the potential for coordination and providing class insights, concerns include social comparison effects \cite{fleur_SocialComparisonLearning_23} and increased student workload. Balancing AI requires an ethical, human-centric approach that enhances learning without diminishing instructor-student interactions \cite{khosravi_XAI_22, alfredo_HAI_23}.

Future work will involve authentic deployments and studying continuous use of \appAbb, instructor interaction, and learning impacts, as well as refining RAG retrieval \cite{gao_rag_24} for better curriculum alignment. Additionally, the current need for manual instructor validation of AI outputs limits scalability, hence, future iterations could use adaptive algorithms learning from feedback to reduce manual review.

    \bibliographystyle{unsrtnat}
    \bibliography{bibliography}

    \end{document}